\documentclass{vldb}
\usepackage{balance}  

\usepackage{nopageno}

\usepackage{fancyhdr}
\usepackage{amsfonts}
\usepackage{bm}

\usepackage{algorithm}
\usepackage{algorithmic}
\usepackage{amssymb}
\usepackage{graphicx}
\usepackage{latexsym}
\usepackage{subfigure}
\usepackage{wrapfig}
\usepackage{amsmath}
\usepackage{nccmath}
\usepackage{bm}
\usepackage{mathtools}

\DeclarePairedDelimiter{\norm}{\lVert}{\rVert}

\makeatletter
\@addtoreset{equation}{section}
\makeatother

\newcommand{\kTO}       {{\bf to }}

\newcommand{\kETAL}    {{\em et al. }}

\newcommand{\mM}        {{\mathcal{M}}}
\newcommand{\mP}        {{\mathcal{P}}}
\newcommand{\mN}        {{\mathcal{N}}}
\newcommand{\mA}        {{\mathcal{A}}}
\newcommand{\mS}        {{\mathcal{S}}}
\newcommand{\mR}        {{\mathcal{R}}}

\newcommand{\bX}        {\textbf{X}}
\newcommand{\ba}        {\textbf{a}}
\newcommand{\bw}        {\textbf{w}}
\newcommand{\bs}        {\textbf{s}}
\newcommand{\bB}        {\textbf{B}}
\newcommand{\bA}        {\textbf{A}}
\newcommand{\bI}        {\textbf{I}}

\newcommand{\sharedaffiliation}[0]{
    \end{tabular}
    \begin{tabular}{c}}

\graphicspath{{./figs/}}

\DeclareMathOperator*{\argmax}{argmax}

\begin{document}


\title{Model-Free Control for Distributed Stream Data Processing using Deep Reinforcement Learning}

\numberofauthors{1}
\author{
      \alignauthor Teng Li, Zhiyuan Xu, Jian Tang and Yanzhi Wang\\
      \email{\{tli01, zxu105, jtang02, ywang393\}@syr.edu}
      \sharedaffiliation
      \affaddr{Department of Electrical Engineering and Computer Science}  \\
      \affaddr{Syracuse University}   \\
      \affaddr{Syracuse, NY 13244}
      \titlenote{Accepted by VLDB 2018}
}

\maketitle
\thispagestyle{fancy}

\begin{abstract}
In this paper, we focus on general-purpose \emph{Distributed Stream Data Processing Systems (DSDPSs)},
which deal with processing of unbounded streams of continuous data at scale distributedly in real or near-real time.
A fundamental problem in a DSDPS is the scheduling problem (i.e., assigning workload to workers/machines)
with the objective of minimizing average end-to-end tuple processing time.
A widely-used solution is to distribute workload evenly over machines in the cluster in a round-robin manner,
which is obviously not efficient due to lack of consideration for communication delay.
Model-based approaches (such as queueing theory) do not work well either due to the high complexity of the system environment.

We aim to develop a novel model-free approach that can learn to well control a DSDPS from its experience
rather than accurate and mathematically solvable system models, just as a human learns a skill (such as cooking, driving, swimming, etc).
Specifically, we, for the first time, propose to leverage emerging Deep Reinforcement Learning (DRL) for enabling model-free
control in DSDPSs; and present design, implementation and evaluation of a novel and highly effective DRL-based
control framework, which minimizes average end-to-end tuple processing time by jointly learning the system
environment via collecting very limited runtime statistics data and making decisions under the guidance of
powerful Deep Neural Networks (DNNs).
To validate and evaluate the proposed framework, we implemented it based on a widely-used DSDPS, Apache Storm,
and tested it with three representative applications: continuous queries, log stream processing and word count (stream version).
Extensive experimental results show 1) Compared to Storm's default scheduler and the state-of-the-art model-based method,
the proposed framework reduces average tuple processing by $33.5\%$ and $14.0\%$ respectively on average.
2) The proposed framework can quickly reach a good scheduling solution during online learning, which justifies
its practicability for online control in DSDPSs.
\end{abstract}


\section{Introduction}
\label{Sec:Intro}
%
In this paper, we focus on general-purpose \emph{Distributed Stream Data Processing Systems (DSDPSs)}
(such as Apache Storm~\cite{Storm} and Google's MillWheel~\cite{Akidau13}), which deal with processing of
unbounded streams of continuous data at scale distributedly in real or near-real time.
Their programming models and runtime systems are quite different
from those of MapReduce-based batch processing systems, such as Hadoop ~\cite{Hadoop} and Spark~\cite{Spark},
which usually handle static big data in an offline manner.
To fulfill the real or near-real time online processing requirements, \emph{average end-to-end tuple processing
time (or simply average tuple processing time~\cite{Xu14})} is the most important performance metric for a DSDPS.

A fundamental problem in a DSDPS is the scheduling problem (i.e., assigning workload to workers/machines)
with the objective of minimizing average tuple processing time.
A widely-used solution is to distribute workload over machines in the cluster in a round-robin manner~\cite{Storm},
which is obviously not efficient due to lack of consideration for communication delay among processes/machines.
We may be able to better solve this problem if we can accurately model the correlation between a solution and its objective value,
i.e., predict/estimate average tuple processing time for a given scheduling solution. However, this is very hard and has not yet been well studied
in the context of DSDPS.
In distributed batch processing systems (such as MapReduce-based systems), an individual
task's completion time can be well estimated~\cite{Zaharia08}, and the scheduling problem can be well formulated into a
Mixed Integer Linear Programming (MILP) problem with the objective of minimizing the make-span of a job~\cite{Chen12,Zhu14}.
Then it can be tackled by optimally solving the MILP problem or using a fast polynomial-time approximation/heuristic algorithm.
However, in a DSDPS, a task or an application never ends unless it is terminated by its user.
A scheduling solution makes a significant impact on the average tuple processing time. But their relationship
is very subtle and complicated. It does not even seem possible to have a mathematical programming formulation for
the scheduling problem if its objective is to directly minimize average tuple processing time.

Queueing theory has been employed to model distributed stream database systems~\cite{Nicola00}.
However, it does not work for DSDPSs due to the following
reasons: 1) The queueing theory can only provide accurate estimations for queueing delay under a few strong assumptions
(e.g, tuple arrivals follow a Poisson distribution, etc), which, however, may not hold in a complex DSDPS.
2) In the queueing theory, many problems in a queueing network (rather than a single queue)
remain open problems, while a DSDPS represents a fairly complicated multi-point to multi-point
queueing network where tuples from a queue may be distributed to multiple downstream queues, and a queue may
receive tuples from multiple different upstream queues.
In addition, a model-based scheduling framework has been proposed in a very recent work~\cite{Li16},
which employs a model to estimate (end-to-end) tuple processing time by combing delay at each component (including
processing time at each Processing Unit (PU) and communication delay between two PUs)
predicted by a supervised learning method, Support Vector Regression (SVR)~\cite{Drucker96}. However, this model-based approach suffers
from two problems: 1) In a very complicated distributed computing environment such as a DSDPS, (end-to-end) tuple processing time (i.e., end-to-end delay)
may be caused by many factors, which are not fully captured by the proposed model.
2) Prediction for each individual component may not be accurate.
3) A large amount of high-dimensional statistics data need to be collected to build and update the model, which leads to
high overhead.

Hence, \emph{we aim to develop a novel model-free approach that can learn to well control a DSDPS
from its experience rather than accurate and mathematically solvable system models,
just as a human learns a skill (such as cooking, driving, swimming, etc).}
Recent breakthrough of \emph{Deep Reinforcement Learning (DRL)}~\cite{Mnih15} provides a promising technique
for enabling effective \emph{model-free} control. \emph{DRL}~\cite{Mnih15} (originally developed by a startup DeepMind)
enables computers to learn to play games, including Atari 2600 video games and one of the most complicated games, Go (AlphaGo~\cite{Silver16}),
and beat the best human players.
Even though DRL has made tremendous successes on game-playing that usually has a limited action space (e.g., moving up/down/left/right),
it has not yet been investigated how DRL can be leveraged for control problems in complex distributed computing systems, such as DSDPSs,
which usually have sophisticated state and huge action spaces.

%
We believe DRL is especially promising for control in DSDPSs because:
1) It has advantages over other dynamic system control techniques such as model-based predictive control
in that the former is model-free and does not rely on accurate and mathematically solvable system models (such as queueing models),
thereby enhancing the applicability in complex systems with randomized behaviors.
2) it is capable of handling a sophisticated state space (such as AlphaGo~\cite{Silver16}),
which is more advantageous over traditional Reinforcement Learning (RL)~\cite{Sutton98}.
3) It is able to deal with time-variant environments such as varying system states and user demands.
%
%
%
However, direct application of the basic DRL technique, such as Deep Q Network (DQN) based DRL proposed in the pioneering work~\cite{Mnih15}, may not work well here
since it is only capable of handling control problems with a limited action space but the control problems (such as scheduling)
in a DSDPS usually have a sophisticated state space, and a huge action (worker threads, worker processes, virtual/physical machines,
and their combinations) space (See Section~\ref{Sec:DRL} for greater details). Moreover, the existing DRL methods~\cite{Mnih15} usually
need to collect big data (e.g., lots of images for the game-playing applications) for learning, which are additional overhead and burden for an online system.
Our goal is to develop a method that only needs to collect very limited statistics data during runtime.

In this paper, we aim to develop a novel and highly effective DRL-based model-free control framework for a DSDPS to minimize average data processing time
by jointly learning the system environment with very limited runtime statistics data and making scheduling decisions under the guidance of powerful \emph{Deep Neural Networks (DNNs)}.
We summarize our contributions in the following:

\begin{itemize}

\item We show that direct application of DQN-based DRL to scheduling in DSDPSs does not work well.

\item We are the first to present a highly effective and practical DRL-based model-free control framework for
\newline scheduling in DSDPSs.

\item We show via extensive experiments with three representative Stream Data Processing (SDP) applications
that the proposed framework outperforms the current practice and the state-of-the-art.

\end{itemize}

\emph{To the best of our knowledge, we are the first to leverage DRL for enabling model-free control in DSDPSs.
We aim to promote a simple and practical model-free approach based on emerging DRL, which, we believe,
can be extended or modified to better control many other complex distributed computing systems.}

The rest of the paper is organized as follows:
We give a brief background introduction in Section~\ref{Sec:Background}.
We then present design and implementation details of the proposed
framework in Section~\ref{Sec:Design}.
Experimental settings are described, and experimental results are presented and
analyzed in Section~\ref{Sec:Eva}.
We discuss related work in Section~\ref{Sec:Related} and conclude the
paper in Section~\ref{Sec:Conclusions}.

\section{Background}
\label{Sec:Background}

In this section, we provide a brief background introduction to DSDPS,
Storm and DRL.

\subsection{Distributed Stream Data Processing System (DSDPS)}
\label{Sec:BG:DSDPS}

In a DSDPS, a stream is an unbounded sequence of tuples.
A data source reads data from external source(s) and emits streams into the system.
A Processing Unit (PU) consumes tuples from data sources
or other PUs, and processes them using code provided by a user.
After that, it can pass it to other PUs for further processing.

A DSDPS usually uses two levels of abstractions
(logical and physical) to express parallelism.
In the logical layer, an application is usually modeled as a directed graph,
in which each vertex corresponds to a data source or a PU, and direct
edges show how data tuples are routed among data sources/PUs.
A task is an instance of a data source or PU, and each data source or PU
can be executed as many parallel tasks on a cluster of machines.
In the physical layer, a DSDPS usually includes a set of virtual or physical machines that actually
process incoming data, and a master serving as the central control unit,
which distributes user code around the cluster, scheduling tasks, and monitoring them for failures.
At runtime, an application graph is executed on multiple
worker processes running on multiple (physical or virtual) machines.
Each machine is usually configured to have multiple slots.
The number of slots indicates the number of worker processes that can be
run on this machine, and can be pre-configured by the
cluster operator based on hardware constraints (such as the
number of CPU cores).
Each worker process occupies a slot, which uses one or multiple threads
to actually process data tuples using user code.
Normally, a task is mapped to a thread at runtime (even it does not have to be this way).
Each machine also runs a daemon that listens for any work assigned to it by the master.
In a DSDPS, a scheduling solution specifies how to assign threads to processes
and machines. Many DSDPSs include a
default scheduler but allow it to be replaced by a custom scheduler.
The default scheduler usually uses a simple scheduling solution,
which assigns threads to pre-configured processes  and then assigns those processes
to machines both in a round-robin manner. This solution leads to almost even
distribution of workload over available machines in the cluster.
In addition, a DSDPS usually supports several ways for grouping, which defines how to distribute
tuples among tasks. Typical grouping policies include: fields grouping (based on a key),
shuffle grouping (random), all grouping (one-to-all) and global grouping (all-to-one).
%


When the message ID of a tuple coming out of a spout
successfully traverses the whole topology, a special acker is called to inform the originating data source that message
processing is complete. The (end-to-end) tuple processing time is the duration between when
the data source emits the tuple and when it has been acked (fully processed).
Note that we are only interested in processing times of those tuples emitted by data sources
since they reflect the total end-to-end processing delay over the whole application.
To ensure fault tolerance, if a message ID is marked failure due to acknowledgment timeout, data processing
will be recovered by replaying the corresponding data source tuple.
The master monitors heartbeat signals from all worker processes periodically.
It re-schedules them when it discovers a failure.

\subsection{Apache Storm}
\label{Sec:BG:Storm}

Since we implemented the proposed framework based on Apache Storm~\cite{Storm}, we briefly
introduce it here. Apache Storm is an open-source and fault-tolerant DSDPS, which
has an architecture and programming model very similar to what described above, and
has been widely used by quite a few companies and institutes.
In Storm, data source, PU, application graph, master, worker process and worker thread are called spout, bolt,
topology, Nimbus, worker and executor, respectively.
Storm uses \emph{ZooKeeper}~\cite{Zookeeper} as a coordination
service to maintain it's own mutable configuration (such as scheduling solution),
naming, and distributed synchronization among machines.
All configurations stored in ZooKeeper are organized
in a tree structure.
Nimbus (i.e., master) provides interfaces to fetch or update Storm's mutable configurations.
%
%
A Storm topology contains a topology specific configuration, which is loaded before the topology starts and does not change during runtime.
%
\subsection{Deep Reinforcement Learning (DRL)}
\label{Sec:BG:DRL}
A DRL agent can be trained via both offline training and online deep Q-learning~\cite{Mnih15,Silver16}.
It usually adopts a DNN (known as DQN) to derive the correlation between each state-action pair $(\bs,\ba)$ of the system
under control and its \emph{value function} $Q(\bs,\ba)$, which is the expected cumulative (with discounts) reward function
when system starts at state $\bs$ and follows action $\ba$ (and certain policy thereafter). $Q(\bs,\ba)$ is given as:
\begin{equation}
Q(\bs,\ba)=\textbf{E}\Big[\sum_{t=0}^{\infty}\lambda^k r_t(\bs_t,\ba_t)\Big|\bs_0=\bs,\ba_0=\ba\Big],
\end{equation}
where $r_t(\cdot)$ is the reward, and $\lambda<1$ is the discount factor.

The offline training needs to accumulate enough samples of value estimates and the corresponding state-action pair $(\bs,\ba)$ for constructing a sufficiently accurate DNN using either a model-based (a mathematical model) procedure or actual measurement data (model-free)~\cite{Mnih15}.
For example, in game-playing applications \cite{Mnih15}, this procedure includes pre-processing game playing samples,
and obtaining state transition samples and Q-value estimates (e.g., win/lose and/or the score achieved).
The deep Q-learning is adopted for the online learning and dynamic control based on the offline-built DNN.
More specifically, at each decision epoch $t$, the system under control is at a state $\bs_t$.
The DRL agent performs inference using the DNN to select action $\ba_t$, either the one with the highest Q-value estimate, or with a certain degree of randomness using the $\epsilon$-greedy policy~\cite{Restelli15}.
%
%

Using a neural network (or even a DNN) as a function approximator in RL is known to suffer from instability or even divergence.
Hence, \emph{experience replay} and \emph{target network} were introduced in~\cite{Mnih15} to improve stability.
A DRL agent updates the DNN with a mini-batch from the experience replay buffer~\cite{Mnih15}, which
stores state transition samples collected during training.
Compared to using only immediately collected samples, uniformly sampling from the replay buffer allows the DRL agent to
break the correlation between sequential generated samples, and learn from a more independently and identically distributed
past experiences, which is required by most of training algorithms, such as Stochastic Gradient Descent (SGD).
So the use of experience replay buffer can smooth out learning and avoid oscillations or divergence.
Besides, a DRL agent uses a separate target network (with the same structure as the original DNN) to estimate target values for training the DNN.
Its parameters are slowly updated every $C>1$ epochs and are held fixed between individual updates.

The DQN-based DRL only works for control problems with a low-dimensional discrete action space.
Continuous control has often been tackled by the actor-critic-based policy gradient approach~\cite{Silver14}.
%
%
The traditional actor-critic approach can also be extended to use a DNN (such as DQN)
to guide decision making~\cite{Lillicrap16}.
A recent work~\cite{Lillicrap16} from DeepMind introduced an actor-critic method, called
Deep Deterministic Policy Gradient (DDPG), for continuous control. The basic idea is to maintain a
parameterized actor function and a parameterized critic function.
The critic function can be implemented using the above DQN, which returns Q value
for a given state-action pair.
The actor function can also be implemented using a DNN, which specifies the current policy by mapping a state to
a specific action.  Both the experience replay and target network introduced above
can also be integrated to this approach to ensure stability.

\section{Design and Implementation of the Proposed Framework}
\label{Sec:Design}
%
In this section, we present the design and implementation details of the proposed framework.

\subsection{Overview}
\label{Sec:Overview}
\begin{figure}[!ht]
\centering
\hfil
\includegraphics[width=0.45\textwidth]{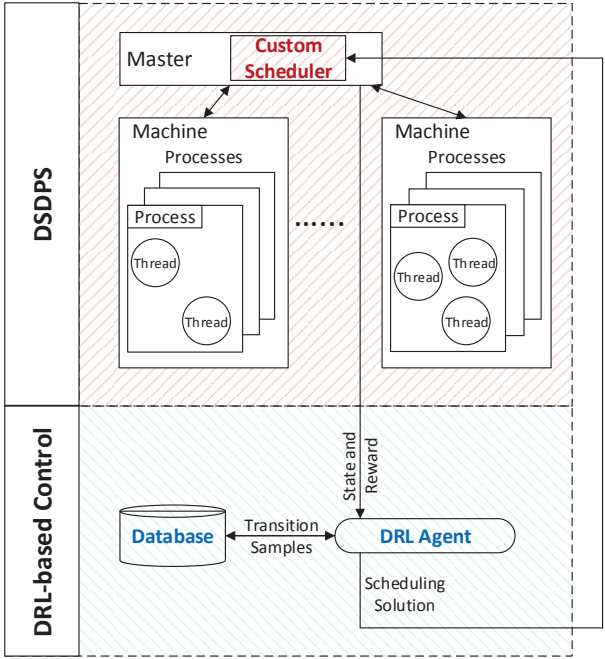}
\caption{The architecture of the proposed DRL-based control framework}
\label{Fig:Architecture}
\end{figure}

We illustrate the proposed framework in Figure~\ref{Fig:Architecture},
which can be viewed to have two parts: DSDPS and DRL-based Control.
The architecture is fairly simple and clean, which consists of the following components:
%
%

\begin{itemize}


\item[1)] \emph{DRL Agent (Section~\ref{Sec:DRL}):} it is the core of the proposed framework, which takes the state as input,
applies a DRL-based method to generating a scheduling solution, and pushes it to the custom scheduler.
%

\item[2)] \emph{Database:} It stores transition samples including state, action and reward information for training (See Section~\ref{Sec:DRL} for details).
%


\item[3)] \emph{Custom Scheduler:}  It deploys the generated scheduling solution on the DSDPS via the master.
%

%

\end{itemize}
%
%
%

Our design leads to the following desirable features:

1) \emph{Model-free Control:} Our design employs a DRL-based method for control, which learns to control a DSDPS
from runtime statistics data without relying on any mathematically solvable system model.

2) \emph{Highly Effective Control:} The proposed DRL-based control is guided by DNNs, aiming to directly minimize average tuple processing time.
Note that the current practise evenly distributes workload over machines; and some existing methods aim to achieve an indirect goal, (e.g., minimizing inter-machine traffic load~\cite{Xu14}),
with the hope that it can lead to minimum average tuple processing time. These solutions are obviously less convincing and
effective than the proposed approach.

3) \emph{Low Control Overhead:} The proposed framework only needs to collect very limited statistics data,
i.e., just the average tuple processing time, during runtime for offline training and
online learning (see explanations in Section~\ref{Sec:DRL}), which leads to low control overhead.

4) \emph{Hot Swapping of Control Algorithms:} The core component of the proposed framework, DRL agent, is external to the DSDPS, which
ensures minimum modifications to the DSDPS, and more importantly, makes it possible to replace it or its algorithm at runtime without
shutting down the DSDPS.

5) \emph{Transparent to DSDSP users:} The proposed framework is completely transparent to DSDSP users, i.e., a user does not have to make any change to
his/her code in order to run his/her application on the new DSDPS with the proposed framework.

We implemented the proposed framework based on Apache Storm~\cite{Storm}.
In our implementation, the custom scheduler runs within Nimbus, which has access to various information
regarding executors, supervisors and slots. A socket is implemented for communications between the custom scheduler and the DRL agent.
When an action is generated by the DRL agent, it is translated to a Storm-recognizable scheduling solution and
pushed to the custom scheduler.
Upon receiving a scheduling solution, the custom scheduler first frees the executors that need to be re-assigned and
then adds them to the slots of target machines.
Note that during the deployment of a new scheduling solution, we try to make a minimal impact to the DSDPS by
only re-assigning those executors whose assignments are different from before while keeping the rest untouched
(instead of deploying the new scheduling solution from scratch by freeing all the executors first and assigning
executors one by one as Storm normally does).
In this way, we can reduce overhead and make the system to re-stabilize quickly.
%
%
In addition, to make sure of accurate data collection, after a scheduling solution is applied, the proposed framework waits for a few minutes
until the system re-stabilizes to collect the average tuple processing time and takes the average of $5$ consecutive measurements
with a $10$-second interval.
%

\subsection{DRL-based Control}
\label{Sec:DRL}
%
In this section, we present the proposed DRL-based control,
which targets at minimizing the end-to-end average tuple processing time via scheduling.
%

Given a set of machines $\mM$, a set of processes $\mP$, and a set of threads $\mN$, a scheduling problem
in a DSDPS is to assign each thread to a process of a machine, i.e., to find two mappings: $\mN \mapsto \mP$
and $\mP \mapsto \mM$. It has been shown~\cite{Xu14} that assigning threads from an application (which
usually exchange data quite often) to more than one processes on a machine introduces inter-process traffic,
which leads to serious performance degradation. Hence, similar as in~\cite{Xu14,Li16}, our design ensures that on every machine,
threads from the same application are assigned to only one process.
Therefore the above two mappings can be merged into just one mapping:
$\mN \mapsto \mM$, i.e., to assign each thread to a machine.
Let a scheduling solution be $\bX = <x_{ij}>, i \in \{ 1, \cdots, N\}, j \in \{ 1, \cdots, M\}$, where $x_{ij} = 1$
if thread $i$ is assigned to machine $j$; and $N$ and $M$ are the numbers of threads and machines respectively.
Different scheduling solutions lead to different tuple processing and transfer delays at/between tasks at runtime
thus different end-to-end tuple processing times~\cite{Li16}.
We aim to find a scheduling solution that minimizes the average end-to-end tuple processing time.

We have described how DRL basically works in Section~\ref{Sec:BG:DRL}. Here, we discuss how to apply
DRL to solving the above scheduling problem in a DSDPS. We first define
the state space, action space and reward.

\emph{State Space}: A state $\bs = (\bX, \bw)$ consists of two parts: the scheduling solution $\bX$, i.e., the current assignment of executors, and the workload $\bw$, which includes the tuple arrival rate (i.e., the number of tuples per second) of each data source.
The state space is denoted as $\mS$.
Workload is included in the state to achieve better adaptivity and sensitivity to the incoming workload, which has been validated by our experimental results.

%
%

\emph{Action Space}: An action is defined as $\ba = <a_{ij}>, \forall i \in \{ 1, ..., N \}, \forall j \in \{ 1, ..., M\}$, where $\sum_{j=1}^M a_{ij}= 1, \forall i$, and $a_{ij} = 1$ means assigning thread $i$ to machine $j$. The action space $\mathcal{A}$ is the space that contains all feasible actions. Note that the constraints $\sum_{j=1}^M a_{ij}= 1, \forall i$ ensure that each thread $i$ can only be assigned to a single machine, and the size of action space $|\mathcal{A}|=M^N$.
Note that an action can be easily translated to its corresponding scheduling solution.

\emph{Reward}: The reward is simply defined to be the negative average tuple processing time so that the objective of the DRL agent is to maximize the reward.

Note that the design of state space, action space and reward is critical to the success of a DRL method. In our case, the action space
and reward are straightforward. However, there are many different ways for defining the state space because a DSDPS includes various
runtime information (features)~\cite{Li16}, e.g., CPU/memory/network usages of machines, workload at each executor/process/ machine, average tuple
processing delay at each executor/PU, tuple transfer delay between executors/PUs, etc.
We, however, choose a simple and clean way in our design. We tried to add additional system runtime
information into the state but found that it does not necessarily lead to performance improvement.
Different scheduling solutions lead to different values for the above features and eventually different average end-to-end tuple processing times;
%
and the tuple arrival rates reflect the incoming workload. These information turns out to be sufficient for representing runtime system state.
%
%
We observed that based on our design, the proposed DNNs can well model the correlation between the state and the average end-to-end tuple processing time (reward)
after training.
%
%
%


A straightforward way to apply DRL to solving the schedu-ling problem is to
directly use the  DQN-based method proposed in the pioneering work~\cite{Mnih15}.
The DQN-based method uses a value iteration approach, in which the value function $Q = Q(\bs, \ba; \bm{\theta})$
is a parameterized function (with parameters $\bm{\theta}$) that takes state $\bs$ and the action space $\mA$ as input
and return Q value for each action $\ba \in \mA$. Then we can use a greedy method to make an action selection:
%
%
\begin{equation}
\pi_Q (\bs) = \argmax_{\ba \in \mA} Q(\bs, \ba; \bm{\theta})
\end{equation}
If we want to apply the DQN-based method here, we need to restrict the exponentially-large action space $\mA$ described above
to a polynomial-time searchable space. The most natural way to achieve this is to restrict each action to assigning only
one thread to a machine. In this way, the size of the action space can be significantly reduced to $|\mA|= N \times M$,
which obviously can be searched in polynomial time.
Specifically, as mentioned above, in the offline training phase, we can collect enough samples of rewards
and the corresponding state-action pairs for constructing a sufficiently accurate DQN, using a model-free method that
deploys a randomly-generated scheduling solution (state), and collect and record the corresponding
average tuple processing time (reward).
Then in the online learning phase, at each decision epoch $t$, the DRL agent obtains the estimated $Q$ value
from the DQN for each $\ba_t$ with the input of current state $\bs_t$.
Then $\epsilon$-greedy policy~\cite{Restelli15} is applied to select the action $\ba_t$ according to the current state $\bs_t$:
with $(1-\epsilon)$ probability, the action with the highest estimated $Q$ value is chosen, or an action is randomly selected with probability $\epsilon$.
After observing immediate reward $r_t$ and next state $\bs_{t + 1}$, a state transition sample $(\bs_t, \ba_t, r_t, \bs_{t + 1})$ is stored into
the experience replay buffer.
At every decision epoch, the DQN is updated with a mini-batch of collected samples in the experience replay buffer using SGD.
%
%

Although this DQN-based method can provide solutions to the scheduling problem and
does achieve model-free control for DSDPSs, it faces the following issue.
On one hand, as described above, its time complexity grows linearly with $|\mA|$, which demands an action space with a very
limited size. On the other hand, restricting the action space may result in limited exploration of the entire exponentially-large
action space and thus suboptimal or even poor solutions to the scheduling problem, especially for the large cases.
The experimental results in Section~\ref{Sec:Eva} validate this claim.
\subsubsection{The Actor-critic-based Method for Scheduling}
\label{Sec:Actor-Critic}
In this section, we present a method that can better explore the action space while keeping time complexity
at a reasonable level.

\begin{figure}[!ht]
\centering
\hfil
\includegraphics[width=0.45\textwidth]{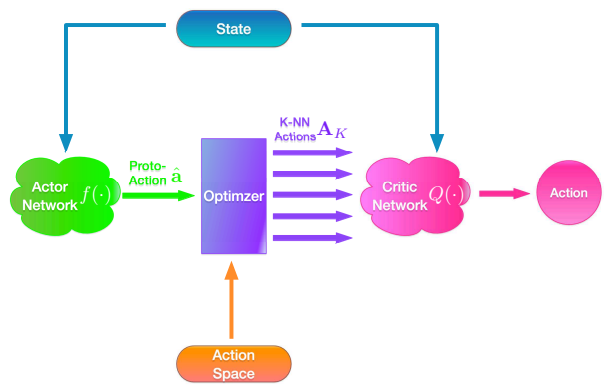}
\caption{The actor-critic-based method}
\label{Fig:Actor-Critic}
\end{figure}

We leverage some advanced RL techniques, including
\newline actor-critic method~\cite{Sutton98,Arnold16} and the deterministic policy gradient~\cite{Silver14},
for solving the scheduling problem. Note that since these techniques only provide general design frameworks, we still need to
come up with a specific solution to our problem studied here.
The basic idea of the proposed scheduling method is illustrated in Figure~\ref{Fig:Actor-Critic}, which includes three major components: 1) an actor network that
takes the state as input and returns a \emph{proto-action} $\hat{\ba}$, 2) an optimizer that finds a set $\bA_K$
of K Nearest Neighbors (K-NN) of $\hat{\ba}$  in the action space, and 3) a critic network that takes the state and
and $\bA_K$ as input and returns Q value for each action $\ba \in \bA_K$. Then an action with the highest Q value can be selected for execution.
The basic design philosophy of this method is similar to that of a rounding algorithm, which finds a continuous solution by
solving a relaxed version of the original integer (i.e., discrete) problem instance and then rounds
the continuous solution to a ``close" feasible integer solution that hopefully offers an objective
value close to the optimal.
Specifically, the actor network $f (\bs; \bm{\theta^{\pi}}) = \hat{\ba}$ is a function parameterized by $\bm{\theta^{\pi}}$ and $f: \mS \mapsto \mathbb{R}^{M^N}$.
$\hat{\ba}$ is returned as a proto-action that takes continuous values so $\hat{\ba} \notin \mA$.
In our design, we use a 2-layer fully-connected feedforward neural network to serve as
the actor network, which includes $64$ and $32$ neurons in the first and second layer respectively and uses the hyperbolic tangent function
$\tanh(\cdot)$ for activation. Note that we chose this activation function because our empirical testing
showed it works better than the other commonly-used activation functions.

The hardest part is to find the K-NN of the proto-action, which has not been well discussed in related work before.
Even though finding K-NN can easily be done in linear time, the input size is $M^N$ here, which could be a huge number even in a small cluster. Hence, enumerating all actions in $\mA$ and doing a linear search to find the K-NN may take an exponentially long time.
We introduce an optimizer, which finds the K-NN by solving a series of Mixed-Integer Quadratic Programming (MIQP)
problems presented in the following:

\noindent MIQP-NN:

\begin{equation}
\begin{split}
&\min_{\ba}: \quad {\norm{\ba - \hat{\ba}}^{2}_{2}}\\
&\mbox{s.t.:} \sum \limits_{j = 1}^{M} a_{ij} = 1, \forall i \in \{ 1, \cdots, N \};\\
&\mbox{\quad} a_{ij} \in \{ 0, 1 \}, \forall i \in \{ 1, \cdots, N \}, \forall j \in \{ 1, \cdots, M \}.
\end{split}
\end{equation}

In this formulation, the objective is to find the action $\ba$ that is the nearest neighbor of the proto-action
$\hat{\ba}$. The constraints ensure that action $\ba$ is a feasible action, i.e., $\ba \in \mA$.
To find the K-NN, the MIQP-NN problem needs to be iteratively solved $K$ times. Each time, one of the KNN of the proto-action
will be returned, the corresponding values $<a_{ij}>$ are fixed, then the MIQP-NN problem is updated and solved again
to obtain the next nearest neighbor until all the K-NN are obtained.
Note that this simple MIQP problem can usually be efficiently solved by a solver as long as the input size is not too large.
In our tests, we found our MIQP-NN problem instances were all solved very quickly (within 10ms on a regular desktop)
by the Gurobi Optimizer~\cite{Gurobi}.
For very large cases, the MIQP-NN problem can be relaxed to a convex programming problem~\cite{Boyd04}
and a rounding algorithm can be used to obtain approximate solutions.

The set $\bA_K$ of K-NN actions are further passed to the critic network to select the action.
The critic network
\newline $Q(\bs, \ba; \bm{\theta^{Q}})$ is a function parameterized by $\bm{\theta^{Q}}$, which returns
Q value for each action $\ba \in \bA_K$, just like the above DQN. The action can then be selected as follows:
\begin{equation}
\pi_Q (\bs) = \argmax_{\ba \in \bA_K} Q(\bs, \ba; \bm{\theta^{Q}}).
\end{equation}
Similar to the actor network $f(\cdot)$, we employ a 2-layer fully-connected feedforward neural network to serve as the critic network,
which includes 64 and 32 neurons in the first and second layer respectively and uses the hyperbolic tangent function $\tanh(\cdot)$
for activation.
Note that the two DNNs (one for actor network and one for critic network) are jointly trained using the collected samples.

\begin{algorithm}[!ht]
\caption{The actor-critic-based method for scheduling}
\label{Alg:Actor-Critic}
\begin{algorithmic}[1]
\STATE Randomly initialize critic network $Q(\cdot)$ and actor network $f(\cdot)$ with weights $\bm{\theta^{Q}}$ and $\bm{\theta^{\pi}}$ respectively;
\STATE Initialize target networks $Q'$ and $f'$ with weights $\bm{\theta^{Q'}} \leftarrow \bm{\theta^{Q}}$, $\bm{\theta^{\pi'}} \leftarrow \bm{\theta^{\pi}}$;
\STATE Initialize experience replay buffer $\bB$; \\
/**Offline Training**/
\STATE Load the historical transition samples into $\bB$, train the actor and critic network offline; \\
/**Online Learning**/
\STATE Initialize a random process $\mathcal{R}$ for exploration;
\STATE Receive a initial observed state $\bs_1$;\\
/**Decision Epoch**/
\FOR{t = 1 \kTO $T$}
	\STATE Derive proto-action $\hat{\ba}$ from the actor network $f(\cdot)$;
	\STATE Apply exploration policy to $\hat{\ba}$: $\mR (\hat{\ba}) = \hat{\ba} + \epsilon \bI$;
	\STATE Find K-NN actions $\bA_K$ of $\hat{\ba}$ by solving a series of MIQP-NN problems (described above);
	\STATE Select action $\ba_t = \argmax_{\ba \in \bA_K} Q(\bs_t, \ba)$;
	\STATE Execute action $\ba_t$ by deploying the corresponding scheduling solution, and observe the reward $r_t$;
	\STATE Store transition sample $(\bs_t, \ba_t, r_t, \bs_{t+1})$ into $\bB$;
	\STATE Sample a random mini-batch of $H$ transition samples $(\bs_i, \ba_i, r_i, \bs_{i+1})$ from $\bB$;
	\STATE $y_i := r_i + \gamma \max_{\ba \in \bA_{i+1, K}} Q'(\bs_{i+1}, \ba)$, $\forall i \in \{1, \cdots, H\}$, where $A_{i+1, K}$ is the set of K-NN of $f'(\bs_{i+1})$;
	\STATE Update the critic network $Q(\cdot)$ by minimizing the loss: \\
	$L(\bm{\theta^Q}) = \frac{1}{H} \sum\limits_{i=1}^{H} [y_i - Q(\bs_i, \ba_i)]^2$;
	\STATE Update the weights $\bm{\theta^{\pi}}$ of actor network $f(\cdot)$ using the sampled gradient: \\
	$\nabla_{\bm{\theta^{\pi}}} f \approx$ $\frac{1}{H} \sum\limits_{i=1}^H \nabla_{\hat{\ba}} Q (\bs, \hat{\ba}) |_{\hat{\ba} = f (\bs_i)} \cdot \nabla_{\bm{\theta^{\pi}}} f (\bs) |_{\bs_i}$; \\
	\STATE Update the corresponding target networks: \\
	$\bm{\theta^{Q'}} := \tau \bm{\theta^{Q}} + (1 - \tau) \bm{\theta^{Q'}}$; \\
	$\bm{\theta^{\pi'}} := \tau \bm{\theta^{\pi}} + (1 - \tau) \bm{\theta^{\pi'}}$ \\
\ENDFOR
\end{algorithmic}
\end{algorithm}

We formally present the actor-critic-based method for sch-eduling as Algorithm~\ref{Alg:Actor-Critic}.
First, the algorithm randomly initializes all the weights $\bm{\theta^{\pi}}$ of actor network $f(\cdot)$ and $\bm{\theta^{Q}}$ of critic network $Q(\cdot)$ (line 1).
If we directly use the actor and critic networks to generate the training target values $<y_i>$ (line 15),
it may suffer from unstable and divergence problems as shown in paper~\cite{Arnold16}.
Thus, similar as in~\cite{Arnold16,Lillicrap16}, we create the target networks to improve training stability.
The target networks are clones of the original actor or critic networks,
but the weights of the target networks $\bm{\theta^{Q'}}$ and $\bm{\theta^{\pi'}}$ are slowly updated,
which is controlled by a parameter $\tau$.
In our implementation, we set $\tau=0.01$.

To robustly train the the actor and critic networks, we adopt the experience replay buffer $\bB$~\cite{Mnih15}.
Instead of training network using the transition sample immediately collected at each decision epoch $t$ (from line 8 to line 12),
we first store the sample into a replay buffer $\bB$, then randomly select a mini-batch of transition samples from $\bB$ to train the actor and critic networks.
Note that since the size of $\bB$ is limited, the oldest sample will be discarded when $\bB$ is full.
%
The sizes of replay buffer and mini-batch were set to $|\bB|=1000$ and $H=32$ respectively in our implementation.

The online exploration policy (line 9) is constructed as $\mR (\hat{\ba}) = \hat{\ba} + \epsilon \bI$, where $\epsilon$ is an adjustable parameter just as the $\epsilon$
in the $\epsilon$-greedy method~\cite{Restelli15}, which determines the probability to add a random noise to the proto-action rather than take the derived action from the actor network.
$\epsilon$ decreases with decision epoch $t$, which means with more training, more derived actions (rather than random ones) will be taken.
In this way, $\epsilon$ can tradeoff exploration and exploitation.
The parameter \textbf{I} is a uniformly distributed random noise, each element of which was set to a random number in $[0, 1]$ in our implementation.

The critic network $Q(\cdot)$ is trained by the mini-batch samples from $\bB$ as mentioned above. For every transition sample $(\bs_i, \ba_i, r_i, \bs_{i+1})$ in the mini-batch, first we obtain the proto-action $\hat{\ba}_{i+1}$ of the next state $\bs_{i+1}$ from the target actor network $f'(\bs_{i+1})$; second, we
find $K$-NN actions $\bA_{i+1, K}$ of the proto-action $\hat{\ba}_{i+1}$ by solving a series of MIQP-NN problems presented above; then we obtain the highest Q-value from the target critic network, $\max_{\ba \in \bA_{i+1, K}} Q'(\bs_{i+1}, \ba)$.
To train critic network $Q(\bs_i, \ba_i)$, the target value $y_i$ for input $\bs_i$ and $\ba_i$ is given by the sum of the immediate reward $r_i$ and the discounted max Q-value (line 15). The discount factor $\gamma=0.99$ in our implementation. A common loss function $L(\cdot)$ is used to train the critic network (line 16).
The actor network $f(\cdot)$ is trained by the deterministic policy gradient method~\cite{Silver14} (line 17).
The gradient is calculated by the chain rule to obtain the expected return from the transition samples in the mini-batch with respect to the weights $\bm{\theta^{\pi}}$ of the actor network.
%

The actor and critic networks can be pre-trained by the historical transition samples, so usually the offline training (line 4) is performed first, which is almost the same as online learning (lines 13--18).
In our implementation, we first collected $10,000$ transition samples with random actions for each experimental setup and then pre-trained the actor and critic networks offline. In this way, we can explore more possible states and actions and significantly speed up online learning.

\section{Performance Evaluation}
\label{Sec:Eva}
%
In this section, we describe experimental setup, followed by experimental results and analysis.

\subsection{Experimental Setup}
\label{Sec:Setup}
We implemented the proposed DRL-based control framework over Apache Storm~\cite{Storm} (obtained from Storm's repository on Apache Software Foundation) and installed the system on top of Ubuntu Linux 12.04. We also used Google's TensorFlow~\cite{TensorFlow} to implement and train the DNNs.
For performance evaluation, we conducted experiments on a cluster in our data center. The Storm cluster consists of 11 IBM blade servers (1 for Nimbus and 10 for worker machines) connected by a 1Gbps network, each with an Intel Xeon Quad-Core 2.0GHz CPU and 4GB memory. Each worker machine was configured to have $10$ slots.

We implemented three popular and representative SDP applications (called topologies in Storm) to test the proposed framework: continuous queries, log stream processing and word count (stream version), which are described in the following.
%
%
%
These applications were also used for performance evaluation in~\cite{Li16} that presented the state-of-the-art model-based approach;
thus using them ensures fair comparisons.

\begin{figure}[!ht]
\centering
\includegraphics[width=0.45\textwidth]{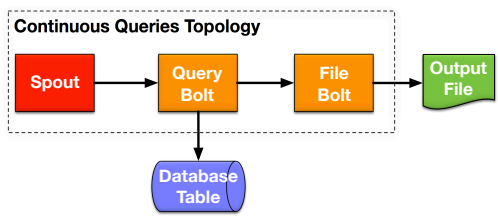}
\caption{Continuous Queries Topology}
\label{Fig:ContinuousQueries}
\end{figure}

\textbf{Continuous Queries Topology (Figure~\ref{Fig:ContinuousQueries})}: This topology represents a popular application on Storm. It is a select query that works by initializing access to a database table created in memory and looping over each row to check if there is a hit~\cite{Bakkum10}. It consists of a spout and two bolts. Randomly generated queries are emitted continuously by the spout and sent to a Query bolt. The database tables are placed in the memory of worker machines. After taking queries from the spout, the Query bolt iterates over the database table to check if there is any matching record. The Query bolt emits the matching record to the last bolt named the File bolt, which writes matching records into a file. In our experiments, a database table with vehicle plates and their owners' information including their names and SSNs was randomly generated. We also randomly generated queries to search the database table for owners of speeding vehicles, while vehicle speeds were randomly generated and attached to every entry.

To perform a comprehensive evaluation, we came up with 3 different setups for this topology: small-scale, medium-scale and large-scale. In the small-scale experiment, a total of 20 executors were created, including 2 spout executors, 9 Query bolt executors and 9 File bolt executors. In the medium-scale experiment, we had 50 executors in total, including 5 spout executors, 25 Query bolt executors and 20 File bolt executors.  For the large-scale experiment, we had a total of 100 executors, including 10 spout executors, 45 Query bolt executors and 45 File bolt executors.

\begin{figure}[!ht]
\centering
\includegraphics[width=0.45\textwidth]{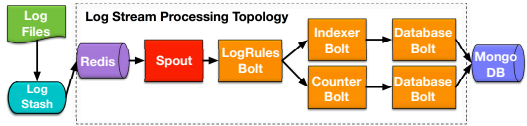}
\caption{Log Stream Processing Topology}
\label{Fig:LogStreamProcessing}
\end{figure}

\textbf{Log Stream Processing Topology (Figure~\ref{Fig:LogStreamProcessing})}: Being one of the most popular applications for Storm, this topology uses an open-source log agent called LogStash~\cite{LogStash} to read data from log files. Log lines are submitted by LogStash as separate JSON values into a Redis~\cite{Redis} queue, which emits the output to the spout. We used Microsoft IIS log files collected from computers at our university as the input data. The LogRules bolt performs rule-based analysis on the log stream, delivering values containing a specific type of log entry instance. The results are simultaneously delivered to two separate bolts: one is the Indexer bolt performing index actions and another is the Counter bolt performing counting actions on the log entries. For the testing purpose, we slightly revised the original topology to include two more bolts, two Database bolts, after the Indexer and Counter bolts respectively. They store the results into separate collections in a Mongo database for verification purpose.

In our experiment, the topology was configured to have a total of 100 executors, including 10 spout executors, 20 LogRules bolt executors, 20 Indexer bolt executors, 20 Counter bolt executors and 15 executors for each Database bolt.

\begin{figure}[!ht]
\centering
\includegraphics[width=0.45\textwidth]{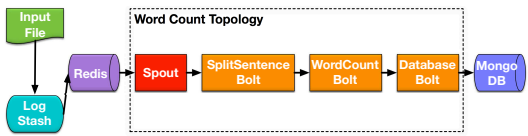}
\caption{Word Count Topology (stream version)}
\label{Fig:WordCount}
\end{figure}

\textbf{Word Count Topology (stream version) (Figure~\ref{Fig:WordCount})}: The original version of the topology is widely known as a classical MapReduce application that counts every word's number of appearances in one or multiple files. The stream version used in our experiments runs a similar routine but with a stream data source. This topology consists of one spout and three bolts with a chain-like structure. LogStash  ~\cite{LogStash} was used to read data from input source files. LogStash submits input file lines as separate JSON values into a Redis~\cite{Redis} queue, which are consumed and emitted into the topology. The text file of Alice's Adventures in Wonderland~\cite{Alice} was used as the input file. When the input file is pushed into the Redis queue, the spout produces a data stream which is first directed to the SplitSentence bolt, which splits each input line into individual words and further sends them to the WordCount bolt. This bolt then counts the number of appearances using fields grouping. The Database bolt finally stores the results into a Mongo database.

In the experiment, the topology was configured to have a total of 100 executors, including 10 spout executors, 30 SplitSentence bolt executors, 30 WordCount executors and 30 Database bolt executors.

\subsection{Experimental Results and Analysis}
\label{Sec:Results}
\begin{figure*}[!ht]
\centerline{
   \subfigure[Small-scale]{\includegraphics[width=0.33\textwidth]{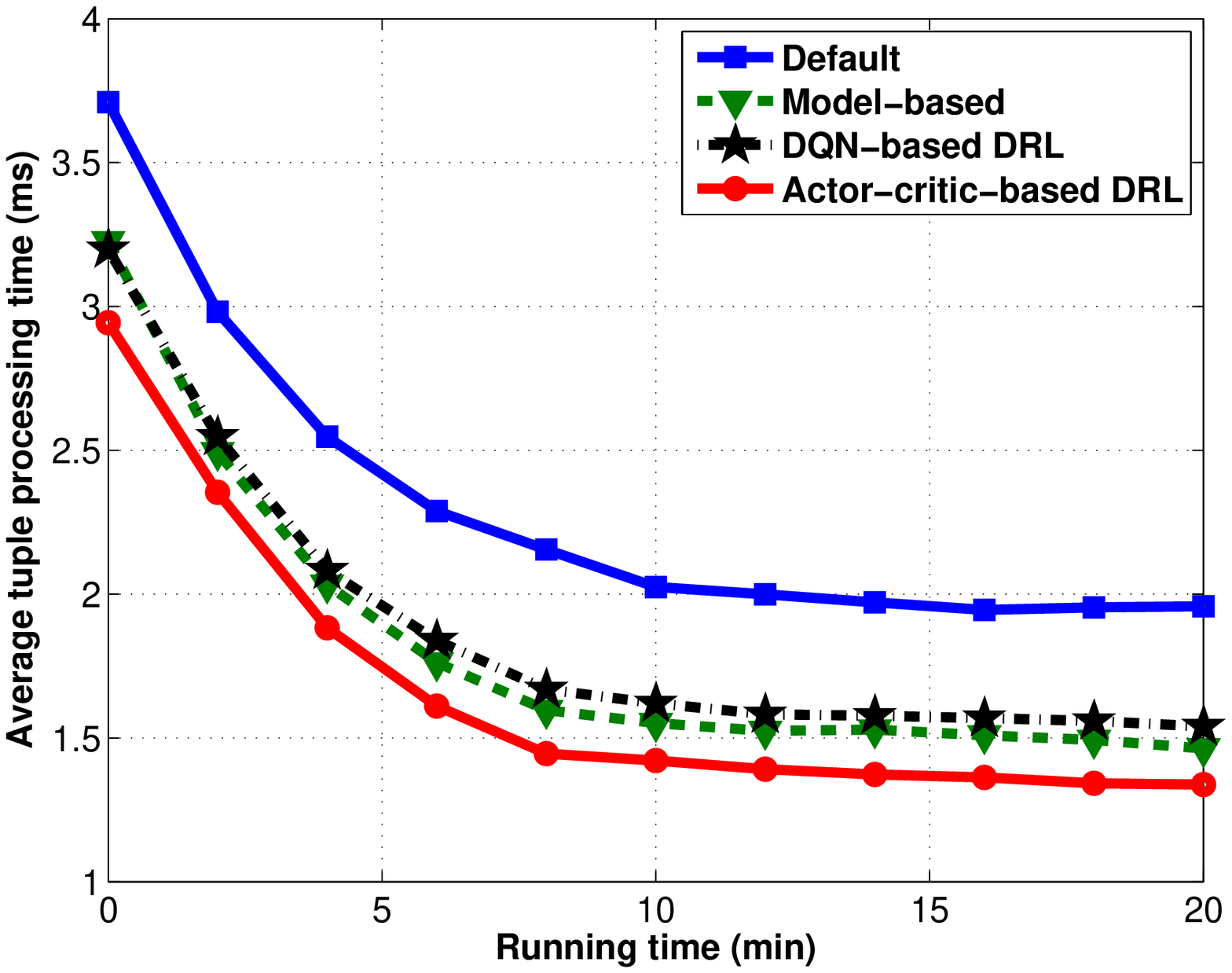}\label{Fig:CQ-Small-T}}
   \hfil
   \subfigure[Medium-scale]{\includegraphics[width=0.33\textwidth]{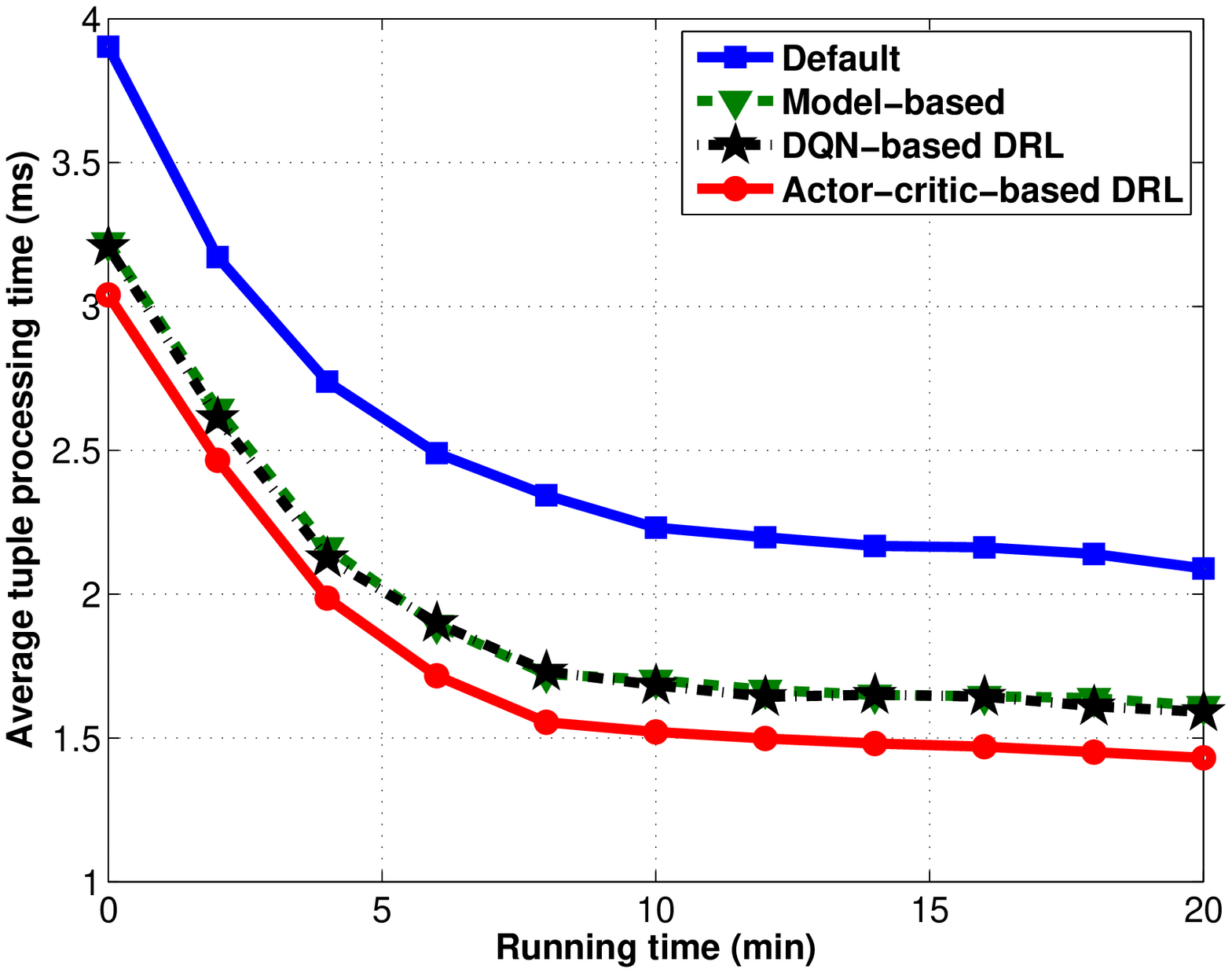}\label{Fig:CQ-Medium-T}}
   \hfil
   \subfigure[Large-scale]{\includegraphics[width=0.33\textwidth]{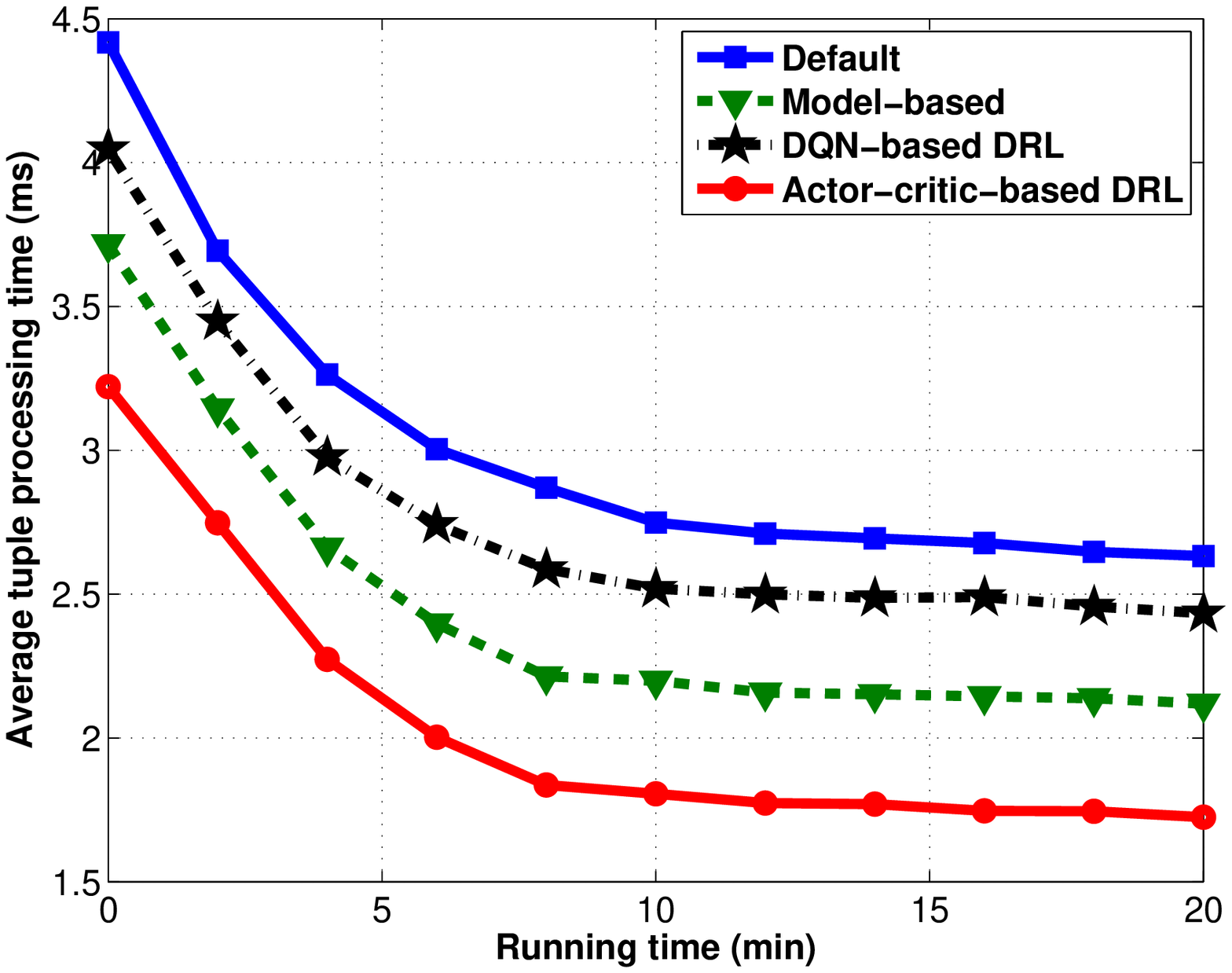}\label{Fig:CQ-Large-T}}
}\caption{Average tuple processing time over the continuous queries topology}
\label{Fig:Average-CQ}
\end{figure*}

\begin{figure}[!ht]
\centering
\includegraphics[width=0.5\textwidth]{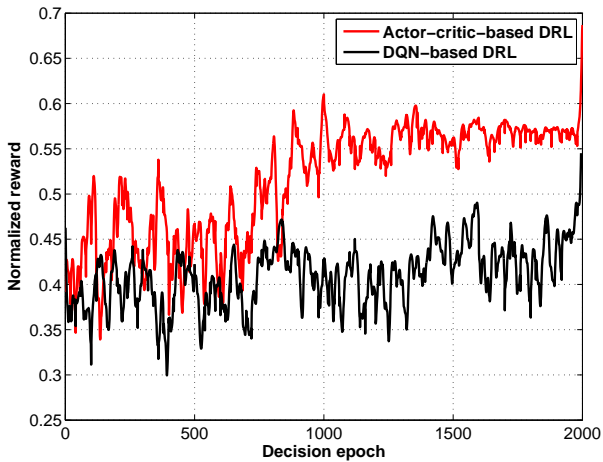}
\caption{Normalized reward over the continuous queries topology (large-scale)}
\label{Fig:CQ-Large-R}
\end{figure}

%

In this section, we present and analyze experimental results. To well justify effectiveness of our design, we compared the proposed DRL-based control framework with the actor-critic-based method (labeled as ``Actor-critic-based
\newline DRL") with the default scheduler of Storm (labeled as ``Default") and the state-of-the-art model-based method proposed in a very recent paper~\cite{Li16} (labeled as ``Model-based") in terms of average (end-to-end) tuple processing time.
Moreover, we included the straightforward DQN-based DRL me-thod (described in Section~\ref{Sec:DRL}) in the comparisons (labeled as ``DQN-based DRL").

\begin{figure}[!ht]
\centering
\includegraphics[width=0.45\textwidth]{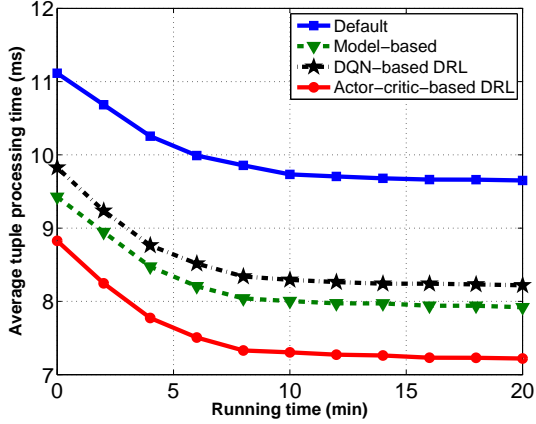}
\caption{Average tuple processing time over the log processing topology (large-scale)}
\label{Fig:Average-LP}
\end{figure}

\begin{figure}[!ht]
\centering
\includegraphics[width=0.45\textwidth]{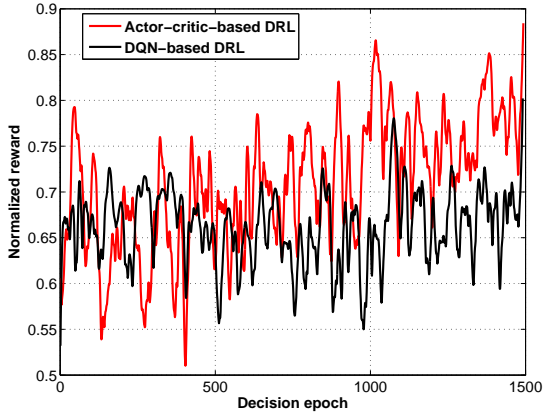}
\caption{Normalized reward over the log processing topology (large-scale)}
\label{Fig:LP-Large-R}
\end{figure}

\begin{figure}[!ht]
\centering
\includegraphics[width=0.45\textwidth]{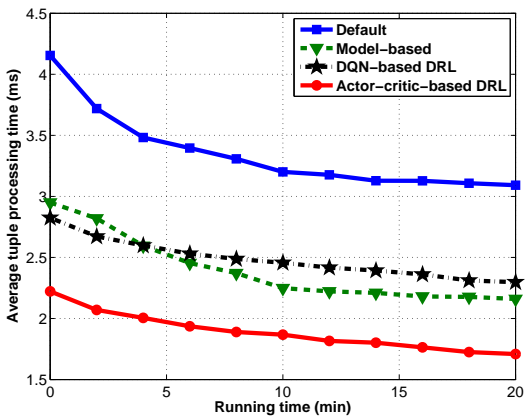}
\caption{Average tuple processing time over the word count topology (large-scale)}
\label{Fig:Average-WC}
\end{figure}

\begin{figure}[!ht]
\centering
\includegraphics[width=0.45\textwidth]{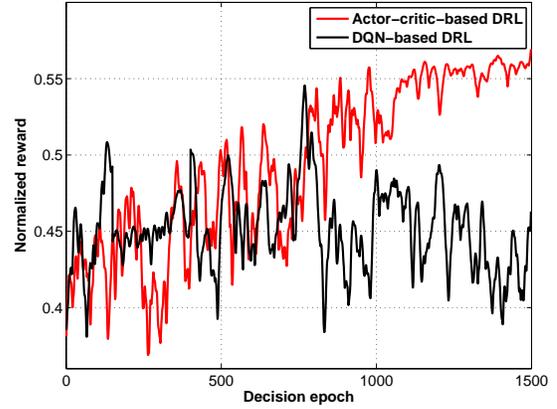}
\caption{Normalized reward over the word count topology (large-scale)}
\label{Fig:WC-Large-R}
\end{figure}

For the proposed actor-critic-based DRL method and the DQN-based DRL method, both the offline training and online learning were performed to train the DNNs to reach
certain scheduling solutions, which were then deployed to the Storm cluster described above.
The figures presented in the following show the average tuple processing time corresponding to the scheduling solutions given by all these methods in the period of 20 minutes. In addition, we show the performance of the two DRL methods over the online learning procedure in terms of the reward. For illustration and comparison purposes, we normalize and smooth the reward values using a commonly-used method $\frac{r - r_{\min}}{r_{\max} - r_{\min}}$ (where $r$ is the actual reward, $r_{\min}$ and $r_{\max}$ are the minimum and maximum rewards during online learning respectively) and the well-known forward-backward filtering algorithm~\cite{Gustafsson96} respectively.
%
%
Note that in Figures~\ref{Fig:Average-CQ}, \ref{Fig:Average-LP} and \ref{Fig:Average-WC}, time 0 is the time when a scheduling solution given by
a \emph{well-trained} DRL agent is deployed in the Storm.
It usually takes a little while (10-20 minutes) for the system to gradually stabilize after a new scheduling solution is deployed.
This process is quite smooth, which has also been shown in~\cite{Li16}.
So these figures do not show the performance of the DRL methods during training processes but the performance of the scheduling solutions
given by well-trained DRL agents. The rewards given by the DRL agents during their online learning processes are shown in Figures~\ref{Fig:CQ-Large-R}, \ref{Fig:LP-Large-R} and \ref{Fig:WC-Large-R}. This process usually involves large fluctuations, which have also been shown in other DRL-related works such as~\cite{Hasselt16,Lillicrap16}.

\textbf{Continuous Queries Topology (Figure~\ref{Fig:ContinuousQueries})}: We present the corresponding experimental results in Figures~\ref{Fig:Average-CQ} and~\ref{Fig:CQ-Large-R}. As mentioned above, we performed experiments on this topology using three setups: small-scale, medium-scale
and large-scale, whose corresponding settings are described in the last subsection.

From Figure~\ref{Fig:Average-CQ}, we can see that for all 3 setups and all the four methods, after a scheduling solution is deployed, the average tuple processing time  decreases and stabilizes at a lower value (compared to the initial one) after a short period of $8-10$ minutes. Specifically, in Figure~\ref{Fig:CQ-Small-T} (small-scale), if the default scheduler is used, it starts at $3.71$ms and stabilizes at $1.96$ms; if the model-based method is employed, it starts at $3.22$ms and stabilizes at $1.46$ms; if the DQN-based DRL method is used, it starts at $3.20$ms and stabilizes at $1.54$ms; and if the actor-critic-based DRL method is applied, it starts at $2.94$ms and stabilizes at $1.33$ms. In this case, the actor-critic-based DRL method reduces the average tuple processing time by $31.4\%$ compared to the default scheduler and by $9.5\%$ compared to the model-based method. The DQN-based DRL method performs slightly worse than the model-based method.

From Figure~\ref{Fig:CQ-Medium-T} (medium-scale), we can see that the average tuple processing times given by all the methods slightly go up. Specifically, if the default scheduler is used, it stabilizes at $2.08$ms; if the model-based method is employed, it stabilizes at $1.61$ms; if the DQN-based DRL method is used, it stabilizes at $1.59$ms; and if the actor-critic-based DRL method is applied, it stabilizes at $1.43$ms. Hence, in this case, the actor-critic-based DRL method achieves a performance improvement of $31.2\%$ over the default scheduler and $11.2\%$ over the model-based method. The performance of DQN-based DRL method is still comparable to the model-based method.

From Figure~\ref{Fig:CQ-Large-T} (large-scale), we can observe that the average tuple processing times given by all the methods increase further
but still stabilize at reasonable values, which essentially shows that the Storm cluster undertakes heavier workload but has not been overloaded in this large-scale case.
Specially, if the default scheduler is used, it stabilizes at $2.64$ms; if the model-based method is employed, it stabilizes at $2.12$ms; if the DQN-based DRL method is used, it stabilizes at $2.45$ms; and if the actor-critic-based DRL method is applied, it stabilizes at $1.72$ms. In this case, the actor-critic-based DRL method achieves a more significant performance improvement of $34.8\%$ over the default scheduler and $18.9\%$ over the model-based method. In addition, the performance of the DQN-based DRL method is noticeably worse than that of the model-based method.
%

In summary, we can make the following observations: 1) The proposed actor-critic-based DRL method consistently outperforms all the other three methods, which well justifies
the effectiveness of the proposed model-free approach for control problems in DSDPSs. 2) The performance improvement (over the default scheduler and the model-based method) offered by the proposed actor-critic-based DRL met-hod become more and more significant with the increase of input size, which shows that the proposed model-free method works even better when the distributed computing environment becomes more and more sophisticated. 3) Direct application of DQN-based DRL method does not work well, especially in the large case. This method lacks a carefully-designed mechanism (such as the proposed MIQP-based mechanism presented in Section~\ref{Sec:Actor-Critic}) that can fully discover the action space and make a wise action selection. Hence, in large cases with huge action spaces, random selection of action may lead to a suboptimal or even poor decision.

We further exploit how the two DRL methods behave during online learning by showing how the normalized reward varies over time within $T=2000$ decision epochs in    Figure~\ref{Fig:CQ-Large-R}. We performed this experiment using the large-scale setup described above. From this figure, we can observe that both methods start from similar initial reward values. The DQN-based DRL method keeps fluctuating during the entire procedure and ends at an average award value of $0.44$ (the average over the last $200$ epochs); while the actor-critic-based DRL method experiences some fluctuations initially, then gradually climbs to a higher value. More importantly, the actor-critic-based DRL method consistently offers higher rewards compared to the DQN-based method during online learning. These results further confirm superiority of the proposed method during online learning. Moreover, we find that even in this large-scale case, the proposed actor-critic-based DRL method can quickly reach a good scheduling solution (whose performance have been discussed above) without going though a really long online learning procedure (e.g., several million epochs) shown in other applications~\cite{Mnih15}. This justifies the practicability of the proposed DRL method for online control in DSPDSs.
%

%
%
%

\textbf{Log Stream Processing Topology (Figure~\ref{Fig:LogStreamProcessing})}:
We performed a large-scale experiment over the log stream processing topology, whose settings have been discussed in the last subsection too.
We show the corresponding results in Figures~\ref{Fig:Average-LP} and \ref{Fig:LP-Large-R}.
This topology is more complicated than the previous continuous queries topology, which leads to
a longer average tuple processing time no matter which method is used.

In Figure~\ref{Fig:Average-LP}, if the default scheduler is used, it stabilizes at $9.61$ms; if the model-based method is employed, it stabilizes at $7.91$ms; if the DQN-based DRL method is used, it stabilizes at $8.19$ms; and if the actor-critic-based DRL method is applied, it stabilizes at $7.20$ms.  As expected, the proposed actor-critic-based DRL method consistently outperforms the other three methods. Specifically, it reduces the average tuple processing time by $25.1\%$ compared to the default scheduler and $9.0\%$ compared to the model-based method. Furthermore, the DQN-based DRL method performs worse than the model-based method, which is consistent with the results related to the continuous queries topology.
Similarly, we show how the normalized reward varies over time within $T=1500$ decision epochs in Figure~\ref{Fig:LP-Large-R}. From this figure, we can make a similar observation that the actor-critic-based DRL method consistently leads to higher rewards compared to the DQN-based method during online learning. Obviously, after a short period of online learning, the proposed actor-critic-based DRL method can reach a good scheduling solution (discussed above) in this topology too.

\textbf{Word Count Topology (stream version) (Figure~\ref{Fig:WordCount})}:  We performed a large-scale experiment over the word count (stream version) topology and
show the corresponding results in Figures~\ref{Fig:Average-WC} and \ref{Fig:WC-Large-R}. Since the complexity of this topology is similar to that of the continuous queries topology, all the four methods give similar average time processing times.

In Figure~\ref{Fig:Average-WC}, if the default scheduler is used,  it stabilizes at $3.10$ms; if the model-based method is employed, it stabilizes at $2.16$ ms; if the DQN-based DRL method is used, it stabilizes at $2.29$ms; and if the actor-critic-based DRL method is applied, it stabilizes at $1.70$ms.
The actor-critic-based DRL method results in a performance improvement of $45.2\%$ over the default scheduler and $21.3\%$ improvement over the model-based method.
The performance of the DQN-based DRL method is still noticeably worse than the model-based method.
We also show the performance of the two DRL methods during online learning in Figure~\ref{Fig:WC-Large-R}, from which we can make observations similar to those related to the first two topologies.


\begin{figure*}[!ht]
\centerline{
   \subfigure[Continuous queries topology]{\includegraphics[width=0.33\textwidth]{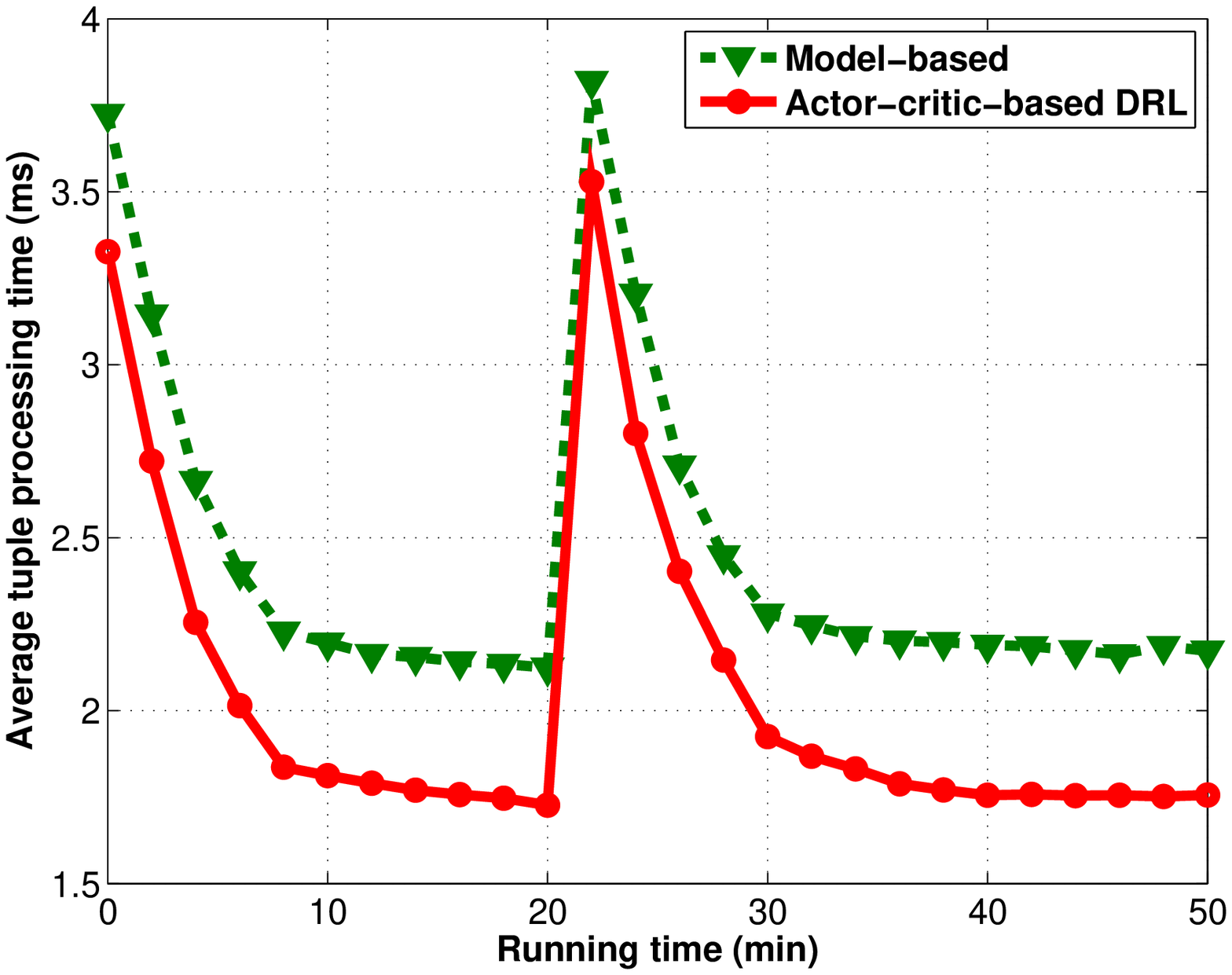}\label{Fig:CQ-Large-F}}
   \hfil
   \subfigure[Log stream processing topology]{\includegraphics[width=0.33\textwidth]{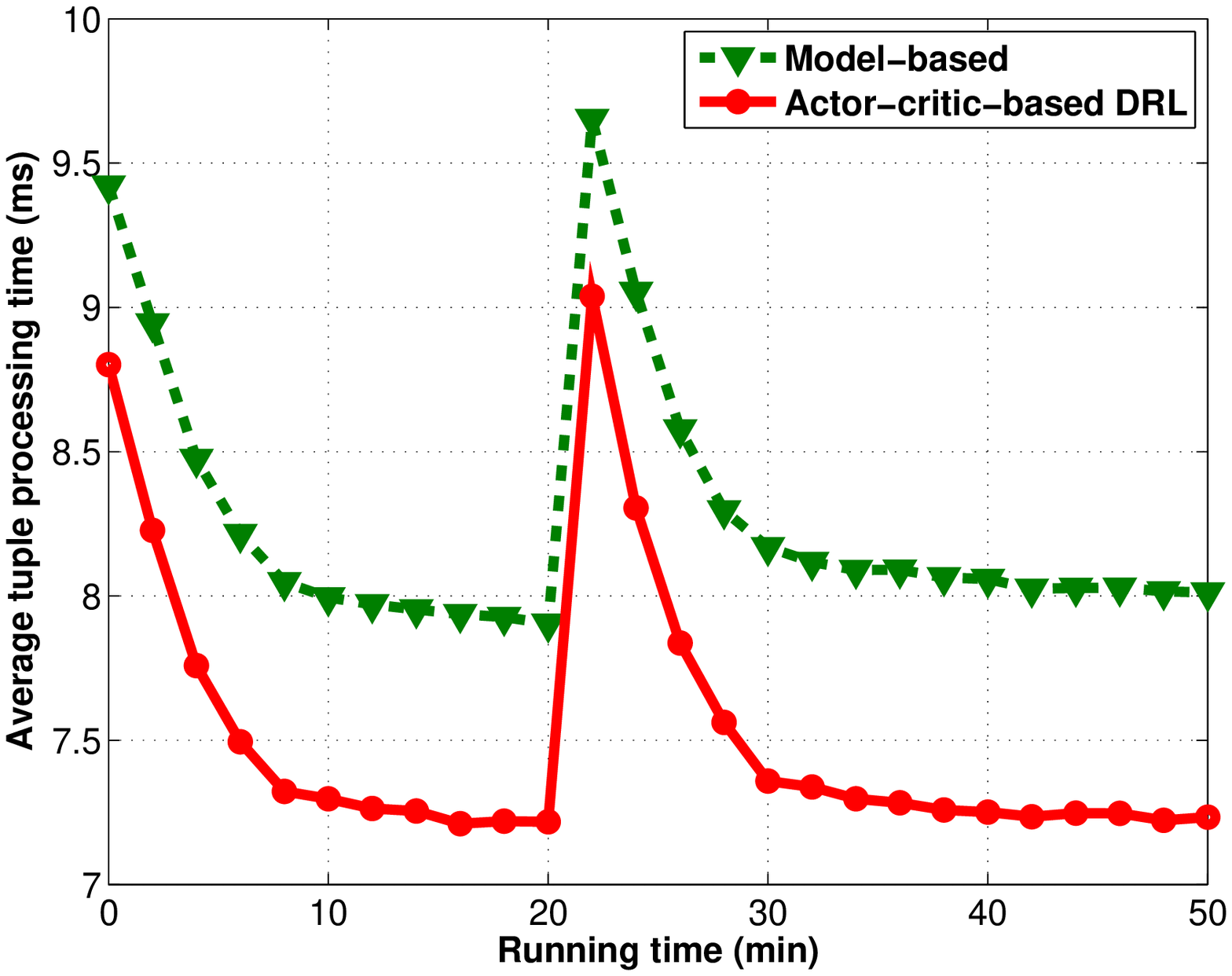}\label{Fig:LP-Large-F}}
   \hfil
   \subfigure[Word count topology]{\includegraphics[width=0.33\textwidth]{./graphs/CQ-Large-Fluctuation}\label{Fig:WC-Large-F}}
}\caption{Average tuple processing time over 3 different topologies (large-scale) under significant workload changes}
\label{Fig:Average-F}
\end{figure*}


In addition, we compared the proposed model-free method with the model-based method under significant workload changes on 3 different topologies (large-scale).
For the continuous queries topology (large-scale), we can see from Figure~\ref{Fig:CQ-Large-F} that when the workload is increased by $50\%$ at 20 minute, the average tuple processing time of the actor-critic-based DRL method rises sharply to a relatively high value then gradually stabilizes at $1.76$ms, while the model-based method rises sharply too and then stabilizes at $2.17$ms. The spikes are caused by the adjustment of the scheduling solution. However, we can observe that once the system stabilizes, the proposed method leads to a very minor increase on average tuple processing time. Hence, it is sensitive to the workload change and can quickly adjust its scheduling solution accordingly to avoid performance degradation. This result well justifies the robustness of the proposed method in a highly dynamic environment with significant workload changes. Moreover, during the whole period, we can see that the proposed method still consistently outperforms the model-based method. We can make similar observations for the other two topologies from Figures~\ref{Fig:LP-Large-F}--\ref{Fig:WC-Large-F}.
\section{Related Work}
\label{Sec:Related}
In this section, we provide a comprehensive review for related
systems and research efforts.

{\bf Distributed Stream Data Processing Systems (DSDPS):}
Recently, a few general-purpose distributed systems
have been developed particularly for SDP in a large
cluster or in the cloud.
Storm~\cite{Storm} is an open-source, distributed and fault-tolerant system that was designed particularly for
processing unbounded streams of data in real or near-real time.
Storm provides a directed graph based model for programming as well as
a mechanism to guarantee message processing.
S4~\cite{S4} is another general-purpose DSDPS that has
a similar programming model and runtime system.
MillWheel~\cite{Akidau13} is a platform designed for building low-latency SDP
applications, which has been used at Google. A user specifies a directed
computation graph and application code for individual nodes, and
the system manages persistent states and continuous flows of
data, all within the envelope of the framework's fault-tolerance
guarantees.
TimeStream~\cite{Qian13} is a distributed system developed by Microsoft specifically for
low-latency continuous processing of stream data in the cloud. It employs
a powerful new abstraction called resilient substitution that
caters to specific needs in this new computation model
to handle failure recovery and dynamic reconfiguration in response
to load changes.
%
Spark Streaming~\cite{SparkStreaming} is a DSDPS that uses the core Spark API and
its fast scheduling capability to perform streaming analytics. It ingests
data in mini-batches and performs RDD transformations on those mini-batches of data.
Other DSDPSs include C-MR~\cite{Backman12}, Apache Flink~\cite{Flink}, Google's FlumeJava~\cite{Chambers10},
M$^3$~\cite{Aly12}, WalmartLab's Muppet~\cite{Lam12},
IBM's System S~\cite{Kumar10} and Apache Samza~\cite{Samza}.

Modeling and scheduling have also been studied in the context of DSDPS and distributed stream database system.
In an early work~\cite{Nicola00}, Nicola and Jarke provided a survey of performance models for distributed and
replicated database systems, especially queueing theory based models.
In~\cite{Jiang03}, scheduling strategies were proposed for a data stream management system,
aiming at minimizing the tuple latency and total memory requirement.
In~\cite{Wei06}, Wei \kETAL proposed a prediction-based Quality-of-Service (QoS) management scheme for periodic queries over dynamic data streams,
featuring query workload estimators that predict the query workload using execution time profiling and input data sampling.
The authors of~\cite{Repantis08} described a decentralized framework for pro-actively predicting and alleviating hot-spots in SDP applications in real-time.
In~\cite{Aniello13}, Aniello \kETAL presented both offline and online schedulers for Storm.
The offline scheduler analyzes the topology structure and makes scheduling decisions;
the online scheduler continuously monitors system performance and reschedules executors at runtime to improve overall performance.
Xu \kETAL~presented T-Storm in~\cite{Xu14}, which is a traffic-aware scheduling framework that minimizes
inter-node and inter-process traffic in Storm while ensuring no worker nodes were overloaded, and enables fine-grained control
over worker node consolidation.
In~\cite{Bedini13}, Bedini \kETAL presented a set of models that characterize a practical DSDPS using the Actor Model theory.
They also presented an experimental validation of the proposed models using Storm.
In~\cite{Bellavista14}, a general and simple technique was presented to design and implement priority-based resource scheduling in flow-graph-based DSDPSs
by allowing application developers to augment flow graphs with priority metadata and by introducing an extensive set of priority schemas.
In~\cite{Khosh15}, a novel technique was proposed for resource usage estimation of SDP workloads in the cloud, which uses mixture density networks, a combined structure of neural networks and mixture models, to estimate the whole spectrum of resource usage as probability density functions.
In a very recent work~\cite{Li16}, Li \kETAL presented a topology-aware method to accurately predict the average tuple
processing time for a given scheduling solution, according to the topology of the application graph and runtime
statistics. For scheduling, they presented an effective algorithm to assign threads to machines under the guidance of the proposed prediction model.

Unlike them, we aim to develop a model-free control framework for scheduling in DSDPSs
to directly minimize the average tuple processing time by leveraging the state-of-the-art DRL techniques,
which has not been done before.

\textbf{Deep Reinforcement Learning (DRL):}
DRL has recently attracted extensive attention from both industry and academia.
In a pioneering work~\cite{Mnih15}, Mnih \kETAL  proposed Deep Q Network (DQN), which can learn successful policies directly
from high dimensional sensory inputs. This work bridges the gap between high-dimensional sensory inputs and actions,
resulting in the first artificial agent that is capable of learning to excel at a diverse array of challenging gaming tasks.
The authors of~\cite{Hasselt16} proposed double Q-learning as a specific adaptation to the DQN, which is introduced in a tabular setting and
can be generalized to work with a large-scale function approximation.
The paper~\cite{Foerster16} considered a problem of multiple agents sensing and acting with the goal of maximizing their shared utility,
based on DQN. The authors designed agents that can learn communication protocols to share information needed for accomplishing tasks.
In order to further extend DRL to address continuous actions and large discrete action spaces,
Duan \kETAL~\cite{Duan16} presented a benchmark suite of control tasks to quantify progress in the domain of continuous control; and Lillicrap \kETAL~\cite{Lillicrap16}
proposed an actor-critic, model-free algorithm based on the policy gradient that can operate over continuous action spaces.
%
%
Gu \kETAL~\cite{Gu16} presented normalized advantage functions to reduce the sample complexity of DRL for continuous tasks.
%
%
Arnold \kETAL~\cite{Arnold16} extended the methods proposed for continuous actions to make decisions within a large discrete action space.
In~\cite{Narasimhan16}, the authors employed a DRL framework to jointly learn state representations and action policies using game rewards as feedback
for text-based games, where the action space contains all possible text descriptions.
%

It remains unknown if and how the emerging DRL can be applied to solving complicated control and resource allocation problems in DSDPSs.

$\\~~~~$In addition, RL/DRL has been applied to control of big data and cloud systems. In~\cite{Naik15}, the authors proposed a novel MapReduce scheduler in heterogeneous environments based on RL, which observes the system state of task execution and suggests speculative re-execution of the slower tasks on other available nodes
in the cluster.
An RL approach was proposed in~\cite{Peng17} to enable automated tuning configuration of MapReduce parameters.
%
%
Liu \kETAL~\cite{Liu17} proposed a novel hierarchical framework for solving the overall resource resource allocation and power management problem in cloud computing systems with DRL.
The control problems in MapReduce and cloud systems are quite different from the problem studied here. Hence, the methods proposed in these related works
cannot be directly applied here.
%
%

\newpage
\section{Conclusions}
\label{Sec:Conclusions}
In this paper, we investigated a novel model-free approach that can learn to well control
a DSDPS from its experience rather than accurate and mathematically solvable system models,
just as a human learns a skill.
We presented design, implementation and evaluation of a novel and highly effective DRL-based
control framework for DSDPSs, which minimizes the average end-to-end tuple processing time.
The proposed framework enables model-free control by jointly learning the system
environment with very limited runtime statistics data and making decisions
under the guidance of two DNNs, an actor network and a critic network.
We implemented it based on Apache Storm, and tested it with three representative applications: continuous queries, log stream processing and word count (stream version).
Extensive experimental results well justified the effectiveness of our design, which showed:
1) The proposed framework achieves a performance improvement
of $33.5\%$ over Storm's default scheduler and $14.0\%$ over the state-of-the-art model-based method on average.
2) The proposed DRL-based framework can quickly reach a good scheduling solution during online learning.
%
\section{Acknowledgement}
\label{Sec:Ack}
This research was supported in part by AFOSR grants FA9550-16-1-0077 and FA9550-16-1-0340.
The information reported here does not reflect the position or the policy of the federal government.



\end{document}